\documentclass[11pt,a4paper]{article} 
\usepackage{jheppub}

\usepackage{amsmath}

\newcommand{\beq}{\begin{equation}}
\newcommand{\eeq}{\end{equation}}

\newcommand{\beqa}{\begin{eqnarray}}
\newcommand{\eeqa}{\end{eqnarray}}

\newcommand{\by}{\begin{eqnarray}}
\newcommand{\ey}{\end{eqnarray}}

\newcommand{\half}{\frac{1}{2}}

\newcommand\fverb{\setbox\fverbbox=\hbox\bgroup\verb}
\newcommand\fverbdo{\egroup\medskip\noindent%
            \fbox{\unhbox\fverbbox}\ }
\newcommand\fverbit{\egroup\item[\fbox{\unhbox\fverbbox}]}
\newbox\fverbbox

\newcommand{\nablaslash}{\not{\hbox{\kern-3pt $\nabla$}}}

\title{Lars Brink: November 12, 1943 - October 29, 2022}

\author{Bengt E.W.~Nilsson and Bj\"orn Jonson
\\
Department of Physics,\\
Chalmers University of Technology,\\
SE-412 96 G\"oteborg, Sweden\\

{\tt {\footnotesize  tfebn@chalmers.se, bjorn.jonson@chalmers.se}}}

\vspace{-5mm}

\abstract{ We give some personal reflections on the person and scientist Lars Brink 
and on some of his scientific achievements. Our relations to Lars are briefly described in
\cite{Nilsson} and \cite{Jonson}, while the  sources relevant for this text are summarised in \cite{sources}.
\begin{figure}[ht!]
\centering
\includegraphics[width=120mm]{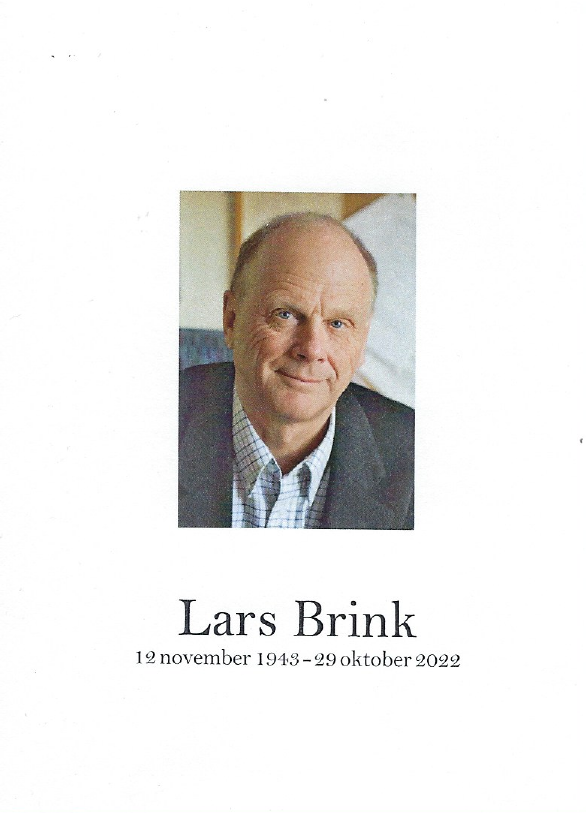}
\end{figure}
}

 \vspace{-5mm}

\toccontinuoustrue
\begin{document}
\maketitle



\section{Lars Brink, the person}

Our friend and colleague Professor  LARS BRINK passed away on October 29,  2022, a short time before his 79th birthday. 
 Between the early 1970s and late 1980s Lars made a number of fundamental contributions to the early developments of dual models and string theory.
One subarea of string theory that he was particularly proud of was maximally supersymmetric Yang-Mills theory and the  many novel results he was
responsible for in works with several different people. 
Some of these  will be briefly described  below. 

Later in his career he became deeply interested in the role of theoretical physics in general
and its growing difficulties to stay relevant in a world focusing more and more on entrepreneurship and applied sciences.
Lars therefore got involved in supporting and helping many scientific institutions outside Sweden, like the Max Planck Institute, the Solvay Institutes in Brussels etc.
Lars was also influential as a  player behind the scene in the  high energy physics community.  The fact that  he knew personally the key people in physics including many 
Nobel laureates in physics gave him ample opportunities to have a say in many global as well as local activities in the community.
In Sweden  he was for many years very active in the main funding agency, NFR later renamed VR. He was also a member of the Royal Academies  
KVA in Stockholm since 1997 and KVVS in G\"oteborg since 2004. In the former case Lars was  also  for several years a member  of the Nobel Committee for Physics and, during the last year 2013, 
also acting as its chairman. Lars was during these years  directly involved in the process of selecting the Nobel Prize laureates in physics.

\begin{figure}[ht!]
\centering
\includegraphics[width=60mm]{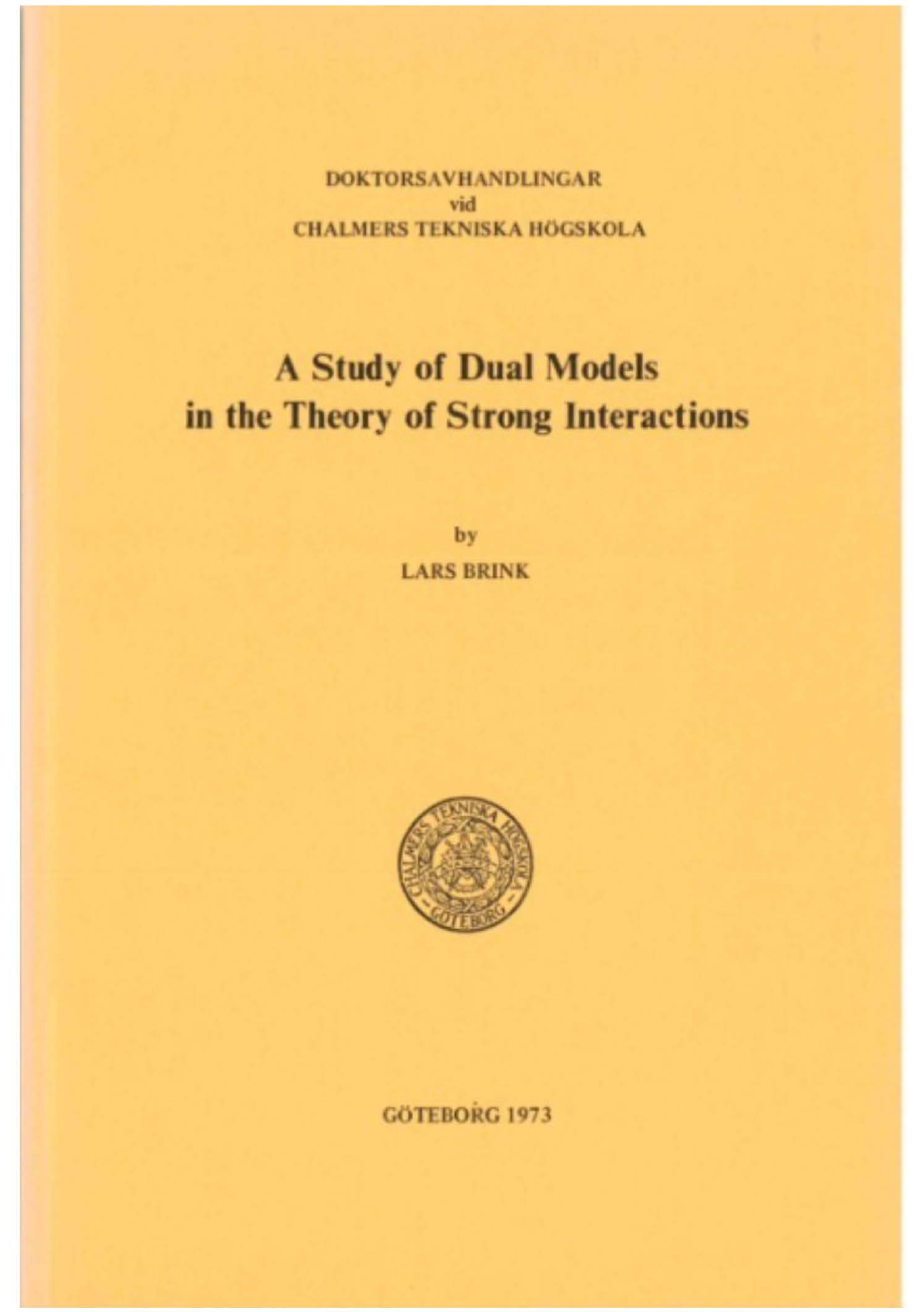}
\caption{Lars Brink's doctoral thesis, December 1973.}
\end{figure}


\section{Lars Brink, the career}

Lars Brink was born on November  12, 1943, in Uddevalla, a small town north of G\"oteborg,  but grew up in G\"oteborg. 
After graduating from high school in 1962  he  continued his physics studies at  
Chalmers Technical University (Chalmers) receiving his master's degree in 1967.  His doctoral studies resulted half-way in a  licentiate degree in 1970 
supervised by  professor Jan S. Nilsson.

In June 1969 Lars attended, as a second year doctoral student, the Lund International Conference on Elementary Particles.
This probably helped him in selecting research topics in the field of high energy particle physics  and led  in 1971 - 1972 to his first  four papers on aspects of scattering amplitudes involving pions and protons
as well as other hadrons.   Judging from the many comments in this direction we have heard over the years, he probably never entirely stopped  thinking  about hadronic physics.

Lars then went to CERN, the European accelerator facility  outside Geneva, 
where he held a position as fellow 1971 - 1973, impressively enough before having received his doctor's 
degree. 
At CERN his career really took off 
and in a short period of time he wrote  most of the 14 papers that in December 1973 made up his doctoral thesis. 
The extremely high quality of the  thesis gave Lars the best possible grade and he was immediately awarded the 
title {\it docent}\footnote{Lars was one of the last students in Sweden to graduate with a doctor's degree. Soon after his thesis defence the transition to the PhD system was completed with a more strict 
study plan and new rules for the  dissertation etc. The grade on the thesis defence was dropped and the respondent  either passed or failed.} . During his time 
at  CERN Lars 
met many of his life-long friends, some of whom would  also become his collaborators. Two  individuals from this period who had a huge impact on his future career
were David Olive and Holger Bech Nielsen, the latter a legendary researcher from  the Niels Bohr Institute in Copenhagen. Another very gifted young researcher
who also participated in these early collaborations  and whom Lars greatly admired was Jo\"el Scherk. Jo\"el tragically died a few years later, in 1980, a fact that  Lars
repeatedly returned to in conversations with his friends.

During the academic year of 1976 - 1977 Lars was a research associate at the California Institute of Technology (Caltech) in Pasadena, Ca, USA.
This turned out to be a very productive time where he wrote his first 
paper on ${\mathcal N}=4$ super-Yang-Mills theory with John Schwarz and Jo\"el Scherk, his most cited 
paper\footnote{L. Brink, J.H. Schwarz and J. Scherk, {\it Supersymmetric Yang-Mills theories}, Nucl. Phys. {\bf B121} (1977), 77-92.}.
He was then given the opportunity to visit Caltech two  months per year for a number 
of years (1977 - 1985) which gave rise to a long-time collaboration with Mike Green and John Schwarz. During these years Lars also met Pierre Ramond which was the start of 
a life-long and very close friendship. These visits resulted in several papers, three of them
together with Murray Gell-Mann, Pierre Ramond and John Schwarz  on supergravity and the relation between the component and superspace descriptions of the ${\mathcal N}=1$ theory. 
Thus Lars worked for many years closely together with the people who later gave one of the most important strings its name, the
Ramond-Neveu-Schwarz-superstring (RNS). Later, Lars received funds from a Swedish  foundation to visit  Chile and 
Claudio Teitelboim's  (now Bunster) institute in Santiago. At one of these visits,  which took place yearly in the period 1986 - 2000, Lars met another life-long friend and collaborator, Marc Henneaux. 
Lars and Marc were, in 2004,  instrumental in the revival of the Solvay Institutes and their famous conference programme.

On February 1, 1986, Lars was appointed professor by the Swedish government, a position that was funded by the Research Council (NFR) and placed
at Chalmers in G\"oteborg. This was one of a very small number of pure research positions at this level that existed  in Sweden. However, Lars 
used the total freedom inherent in this position to get involved in teaching at both lower and higher levels. One reason for this was his
huge interest in interactions with young students and to discuss all kinds of physics with them. The students were so appreciative of his
relation to the students that they gave him a life-long membership and an honorary title, Inspector, in their  student organisation at Chalmers.
During his active time as a supervisor Lars guided 15 students to their PhD degrees, and as an administrator he spent six years as the vice dean
of the Physics Department.

In 1984 Lars accepted the task to work for the Research Council as a member of the physics committee which he continued doing for ten years,
and when the Research Council was reorganised in 2001 he became the  chairman of the physics subcommittee that dealt with the funding of particle physics 
and some other more or less related areas in physics. At the international level, Lars was the Swedish delegate to the board of the Nordic institute
Nordita during the years 1987 to 1995 and acted as its chairman 1990 to 1992.  At the European level Lars was also very active, in particular during the years 
1991 - 1995 and  2000 - 2008 when he led the EU-network "Superstrings" which involved a large number of research groups all over Europe.

Lars became, after some time as an employee at Chalmers, very interested in the conditions under which fundamental research was conducted at Swedish universities.
During the years 2002 - 2008 he was  chairman of the Faculty Senate which is a body whose main role is to guard the academic perspective in the running of the university.
 This gave Lars ample opportunities to act on the threat, as he saw it, from industry and other external forces that tended to inhabit the university boards to a larger and larger  extent
 for each reorganisation imposed by the government. Lars was not afraid of making his opinion on these matters very clear;  among other things, this was obvious from  a sticker he had on his office door saying
{\it Not an entrepreneur and proud of it!}. 
As the chairman of the Faculty Senate  at Chalmers  he also held  a seat on the governing board of the entire university. 
In this capacity he became in 2008   involved in the selection
of a new head of the university (rector) as well as in the reorganisation of the whole university that took place shortly before. In the year 2018 Lars was awarded the Chalmers medal
for his hard work and efforts to improve Chalmers as a technical university.

\begin{figure}[ht!]
\centering
\includegraphics[width=200mm]{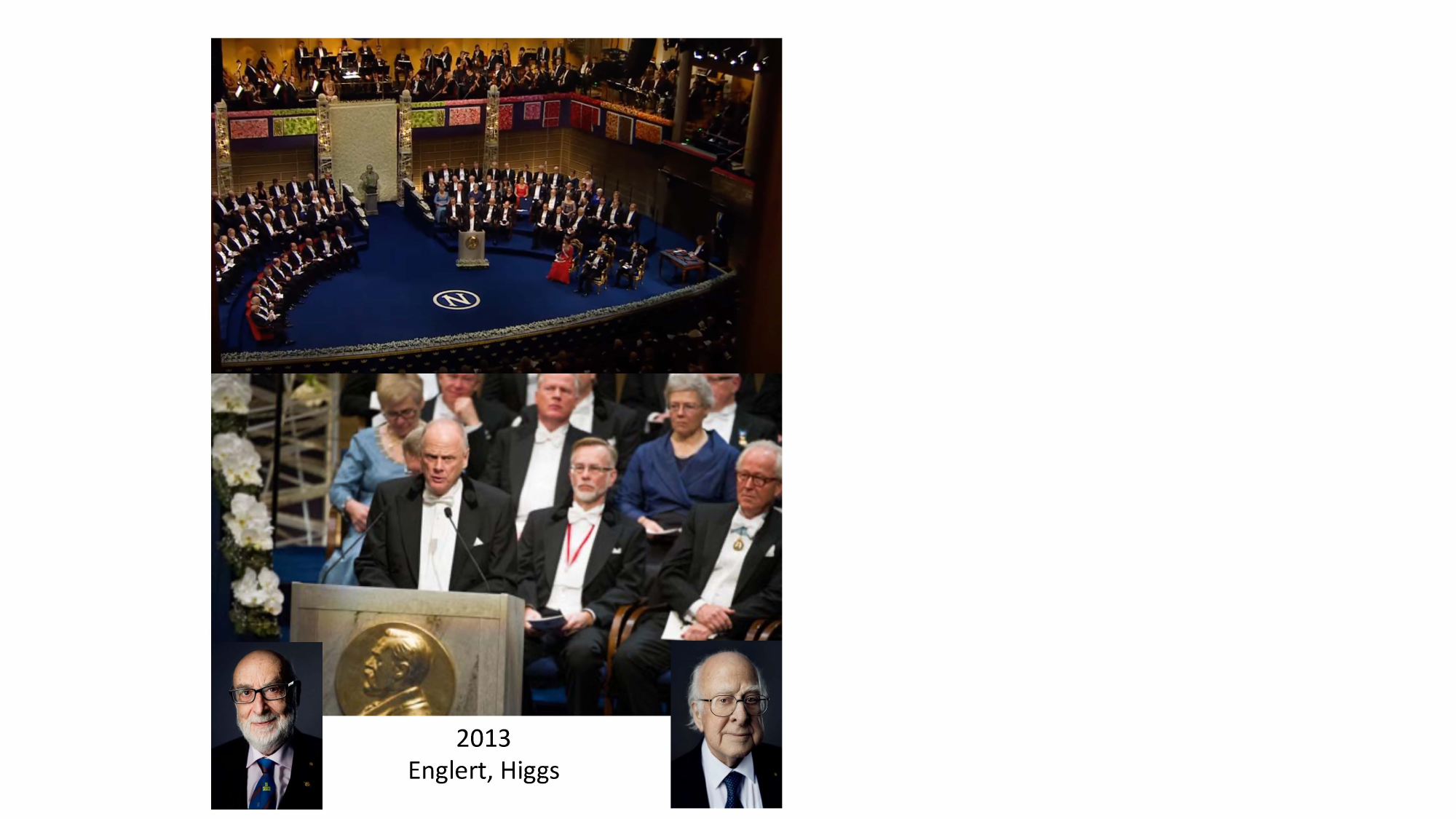}
\caption{Lars Brink's speech at the Nobel ceremony 2013.}
\end{figure}
The next important step in Lars' academic career is connected to him being elected member of KVA, the Swedish Royal Academy of Sciences, in 1997. During the year 2001 
he  was an adjoint member of the Nobel Committee for Physics, which happened again in 2004, a year which ended with him delivering the speech at the Prize ceremony
in the Stockholm Concert Hall. Lars was then a full member of the Physics Committee during 2008 - 2013. 
The 2008 prize in Physics was close to his own research so he once again
delivered  the speech at the Prize ceremony. Since the three laureates at this occasion were all from 
Japan\footnote{The three physics laureates 2008 were M. Kobayashi, T. Maskawa and Y. Nambu.} Lars decided to end his 
speech in their native tongue which he did fluently to the huge surprise of everyone present.
The last year as a member of the Committee he was also its chairman. The third time Lars delivered the speech at the ceremony was in 
2013 (see Figure 2)\footnote{The two physics laureates 2013 were F. Englert and P. Higgs.}.

After his retirement in 2010 Lars continued to work part-time at Chalmers for another three years. Besides his engagement in the Nobel committee he 
was now as emeritus free to accept invitations as guest professor to a number of institutes all around the globe. He visited  for instance the 
Institute of Advanced Studies in Princeton, CERN as a special guest of the General Director of CERN, LMU in Munich and the Institute of Advanced Studies in
Singapore. In the latter case it gave Lars the opportunity to organise a number of conferences and edit volumes of collected works lifting the importance
of the achievements of key people in  high energy physics, some of them Nobel Prize laureates. 
Among the people who were honoured this way were  S. Mandelstam, A. Salam, C.N. Yang and  Y. Nambu. His special relation to  Y. Nambu will be 
commented upon below.

Lars' deep interest in assisting, as an advisor,  different physics institutes  eventually gave him  two honorary doctorates, first at the University of Craiova, 
Romania in 2012, and later, in 2014, at the University of Florida. An  especially important  effort was done in 2004 and subsequent years together with Marc Henneaux at UBL, Brussels, 
in connection with the revival of the Solvay Institutes and their famous conferences. That these conferences have had an enormous impact  historically is without doubt.
One merely needs to recall the one in physics held in 1927. The conference photo (on Wikipedia) is classic: It contains 17 Nobel Prize laureates in physics and chemistry.
Lars' international engagements included being part of several advisory committees, in some cases as chairman, at, e.g., Albert Einstein Institute in Potsdam, Werner-Heisenberg Institute in Munich, 
Centro de Estudios de Scientificos in Valdivia, Chile
and the International Center for Fundamental Physics in Moscow (besides the Solvay Institutes).

When the Covid-19 pandemic started in the spring of 2020 Lars stopped travelling and spent most of his time in his home in Torslanda,  a suburb outside G\"oteborg.
Lars was shortly before the outbreak  informed about his incurable disease.

\section{Lars Brink, the scientist}

Lars gave important contributions to the following  six research topics as will be briefly described below:\\
1. Dual string models:  zero point sum, projection operators  (1971-1973)\\
2. Supergravity and supersymmetry on the worldsheet (1976, 1977, 1987) \\
3. Supergravity and supersymmetry in spacetime (1978 - 1980,  2005, 2008)\\
4. String theory (1976, 1982, 1983) \\
5. Light-cone: N=4 super-Yang-Mills   and N=8 supergravity (1977, 1983, 2008)\\
6. Light-cone: Higher spins (1983,  2000) \\
\\

Lars Brink was one of the real pioneers of the  field of string theory and gave several fundamental contributions to its early developments. As a young 
doctoral student at Chalmers
he attended, as mentioned above,  the Lund International Conference on Elementary Particles held in Lund, Sweden, at the end of June 1969  as a scientific secretary. The proliferation of unstable 
strongly interacting hadronic particles appearing on so called Regge trajectories (see below) 
and emerging ideas about partons, including quarks, were discussed at the meeting. It seems likely that being present at
 this meeting influenced him  when selecting research topics for his graduate studies. Four years later his 
doctoral thesis, defended in December 1973, contained 14 papers some of which  provided key results in dual string models which at the time were theories of hadrons. Lars thanks in his thesis
his supervisor professor Jan S. Nilsson for making his doctoral studies so immensely successful. In the  years
following Lars' dissertation dual models developed into string theory, starting with the realisation in 1974 of 
T. Yoneya\footnote{T. Yoneya, {\it Connection of Dual Models to Electrodynamics and Gravidynamics}, Prog. Theor. Phys. {\bf 51} (1974), 1907 - 1920.} 
and of  two of Lars' close friends Jo\"el Scherk and John Schwarz\footnote{J. Scherk and J.H. Schwarz, {\it Dual Models and Non-hadrons}, Nucl. Phys. {\bf B81} (1974), 118 - 144.} that the spectrum contained
a massless spin 2 particle and hence was actually a model of gravity. Subsequently it was also understood that string theory even was a consistent theory
of quantum gravity. These steps from dual models to string theory, that is from hadron physics  to quantum gravity, turned out to constitute a paradigm shift
 for both high energy particle physics and gravitational research. Lars certainly played a central role in making this transition possible.

Dual models was initially an attempt to understand the ramifications of the huge spectrum of hadronic resonances discovered in the 1950s and 1960s.
These states seem to appear on so called Regge trajectories given by linear relations between spin and mass-squared, $J=\alpha' M^2+\alpha(0)$ where
$\alpha'$ is the Regge slope parameter (about 1 $(GeV)^{-2}$ in hadron physics) and $\alpha(0)$ is the intercept. The key feature of the $2 \rightarrow 2$
scattering amplitude $A(s,t)$ for pions was argued to be that it should satisfy  the  {\it duality} condition  $A(s,t)=A(t,s)$, that is, be symmetric (dual) between the $s$- and $t$-channels.
This way one could perhaps explain how an infinite set of more and more massive hadrons exchanged in the $t$-channel might  avoid breaking unitarity by summing up to poles
corresponding to themselves appearing in the $s$-channel. Thus only one of the channels was needed to get the full result a feature  reminiscent of what happens in string theory.

The first expression discovered that satisfies this requirement was proposed, {\it but not derived},  by Veneziano in 1968\footnote{Actually, the full amplitude for the four-pion scattering is the sum
$A(s,t,u)=A(s,t)+A(s,u)+A(t,u)$.}:
\beq
A_o(s,t)=\frac{\Gamma(-\alpha(s))\Gamma(-\alpha(t))}{\Gamma(-\alpha(s)-\alpha(t))},
\eeq
where $\alpha(s)=\alpha's+\alpha(0)$. In the hadronic context discussed here
the Regge trajectories were assumed to fit the experimental observations. For example, the leading trajectory with at least five data points ($J=1,2,..,5$),
 the so called $\rho$-trajectory, is quite well described by
the linear  $J(M^2)$ relation\footnote{For the numerical values see the textbook discussion in B. Zwiebach, {\it A first course in string theory} (2nd Ed., Cambridge , 2009),  page 527.}
\beq
J=0.87702 (GeV)^{-2}M^2+0.47188.
\eeq

Later a slightly different expression was introduced by Virasoro (1969)  
\beq
A_c(s,t,u)=\frac{1}{stu}\frac{\Gamma(1-\frac{\alpha' s}{4})\Gamma(1-\frac{\alpha' t}{4})\Gamma(1-\frac{\alpha' u}{4})}{\Gamma(1+\frac{\alpha' s}{4})\Gamma(1+\frac{\alpha' t}{4})\Gamma(1+\frac{\alpha' u}{4})}.
\eeq
The index notation on the left-hand side of these two expressions refers to the fact that they later came to be associated with open and closed strings. 
That dual models mathematically could be connected to string-like objects was realised first by Y. Nambu, and within a year later  also by L. Susskind and independently 
H.B. Nielsen. This idea  emerged from the fact that the spin-energy relation of Regge trajectories  could be found also in rotating rigid 
rods\footnote{See also Nambu's own account of the origin of his understanding of these facts in {\it Nambu, A Foreteller of Modern Physics}, Eds. T. Eguchi and M.Y. Han, World Scientific Series in 20th Century Physics, Vol 43 (2014).}. Note that, as presented above,  $A_c(s,t)$ is the form obtained from string theory, i.e., the Regge slope parameter is now related to the Planck length and the intercept
is such that the first excitation is a massless spin 2 state\footnote{This result from string theory seems to have been conjectured first in C. Lovelace, Phys. Lett. {\bf 34B} (1971) 457. This paper may also be  
 the first one to mention the critical dimension $D=26$. }. Clearly, dual string theory could no longer be associated to hadron physics once the intercept was computed mathematically.
It is very interesting that already in his thesis, as a very last comment, Lars compared the Planck's constant $\hbar$ to $\alpha'$ when he asked if the introduction of the latter might in the future turn out 
to be as significant as that of $\hbar$.

The important step of connecting dual models to strings was a result of trying to answer the question of how to generate the two dual amplitudes above from some field theoretic models. The answer
was indicated by the integral  version of the four-point amplitude:
\beq
A_o(s,t)=\frac{\Gamma(-\alpha(s))\Gamma(-\alpha(t))}{\Gamma(-\alpha(s)-\alpha(t))}=\int_0^1 dx x^{-\alpha(s)-1}(1-x)^{-\alpha(t)-1},
\eeq
and entailed a factorisation of the amplitude which Nambu found could be done using an infinite set of harmonic oscillators put together into expressions like
\beq
\label{nambu}
F(p)=e^{i\sqrt{2\alpha'}p^{\mu}\Sigma_{n=1}^{\infty}\frac{a^n_{\mu}x^n}{\sqrt{n}}},\,\,\,G(p)=e^{i\sqrt{2\alpha'}p^{\mu}\Sigma_{n=1}^{\infty}\frac{a^{n\dagger}_{\mu}}{\sqrt{n}}},
\eeq
where the operators commute as $[a_{\mu}^m, a_{\nu}^{n\dagger}]=\eta_{\mu\nu}\delta^{mn}$.
Using the standard Baker-Housdorff-Campbell (BHC) formula applied to operators whose commutator is a c-number one readily finds 
\beq
(1-x)^{-2\alpha'p_2\cdot p_3}=\langle 0|F(p_2)G(p_3)|0 \rangle.
\eeq
A very interesting account of these early developments (taking place September to December 1968) is the one  presented by P. Frampton  in the 2016 memorial volume for 
Y. Nambu that Lars was partly responsible for\footnote{P. Frampton,  in {\it Memorial Volume for Y. Nambu}, Eds. L. Brink, L.N. Chang, M.-Y. Han and K.K. Phua, World Scientific (2016).}. 

Nambu also introduced the important concept of a {\it vertex operator}\footnote{See the contribution of Peter Freund in 
 {\it Memorial Volume for Y. Nambu}, Eds. L. Brink, L.N. Chang, M.-Y. Han and K.K. Phua, World Scientific (2016).} which is an obvious generalisation  of the expressions in  (\ref{nambu}). In modern notation the vertex operator related to the emission of a tachyon state in the closed bosonic string is simply (now with $[\alpha_{\mu}^m, \alpha_{\nu}^{n\dagger}]=m\eta_{\mu\nu}\delta^{mn}$ and similarly for $\bar\alpha_m^{\mu}$)
\beq
V(k^{\mu};(z, \bar z))=:e^{ik_{\mu}X^{\mu}(z, \bar z)}:\equiv e^{ik_{\mu}X^{(+)\mu}(z, \bar z)}e^{ik_{\mu}X^{(-)\mu}(z, \bar z)},
\eeq
where we have made the normal ordering relative a zero-momentum and oscillator vacuum state $|0\rangle_{p} $ explicit: $X^{(+)\mu}(z)|0\rangle_{p} =0$. This follows from the definitions
\beq
X^{(+)\mu}(z,\bar z) \equiv -i\frac{\alpha'}{2}(\log z\bar z) \,p^{\mu} + i\sqrt{\frac{\alpha'}{2}}\Sigma_{n=1}^{\infty}\frac{1}{n}(\alpha_n^{\mu}z^{-n}+\bar\alpha_n^{\mu}{\bar z}^{-n}),
\eeq
\beq
 X^{(-)\mu}(z,\bar z) \equiv x^{\mu} - i\sqrt{\frac{\alpha'}{2}}\Sigma_{n=1}^{\infty}\frac{1}{n}(\alpha^{\mu\dagger}_{n}z^{n}+\bar\alpha^{\mu\dagger}_{n}{\bar z}^{n}),
\eeq
which thus in addition to the oscillators introduced above also contain the (operator) zero modes $x^{\mu}$ and $p^{\mu}$, i.e., the center of mass parameters of the 
string\footnote{These were first introduced in S. Fubini and G. Veneziano, Nuovo Cimento {\bf A 67}, (1970) 29. For an early use of the oscillator expressions see also Phys. Lett. {\bf B29} (1969), 679 by the same authors together with D. Gordon.}.
Here we have expressed it in terms of the complex variables $z=e^{-i(\sigma+i\tau)}$ where $\sigma$ and $\tau$ are the world-sheet coordinates. 
The simplicity of these last two expressions was  noticed by Nambu and can be seen by computing the 
expectation value  of the product of two such operators (today known as an operator product expansion (OPE)):
\beq
{}_x\langle 0| V(k_1^{\mu},z_1)V(k_2^{\mu},z_2)|0 \rangle_p=e^{[ik_{1\mu}X^{(+)\mu}(z_1,\bar z_1),ik_{2\nu}X^{(-)\nu}(z_2,\bar z_2)]}=e^{\alpha'k_1\cdot k_2 \log|z_2-z_1|}=|z_2-z_1|^{\alpha'k_1\cdot k_2},
\eeq
which, again,  is an immediate consequence of the BHC formula. In fact, the contraction, or correlator,
of two string fields $X^{\mu}(\tau,\sigma)$ is given by the normal ordered product $:X^{\mu}(z_2)X^{\mu}(z_1):=-\frac{\alpha'}{2}(\log(z_2-z_1)+\log(\bar z_2-\bar z_1))$ which is the $1+1$ dimensional propagator.
 The N-point  version is then easily obtained and leads to what  is known as the Koba-Nielsen formula. 

It is quite remarkable that Nambu shortly after these discoveries also was the first one to introduce the world-sheet area (in July 1969, see the last two footnotes above), spanned by coordinates $(\tau,\sigma)$, as 
the two-dimensional action principle that generates amplitudes of the kind above, the so-called
Nambu-Goto action\footnote{See Nambu's lecture notes for Copenhagen 1970, and T. Goto, Prog. Theor. Phys {\bf 46} (1971), 1560.}
\beq
S[X^{\mu}]=\frac{1}{4\pi \alpha'}\int d\tau d\sigma \left( \sqrt{(\dot X\cdot X')^2-(\dot X)^2(X')^2} \right),\, \text{where} \,{\dot X}^{\mu}\equiv \frac{\partial X^{\mu}}{\partial \tau},\, X'^{\mu}\equiv \frac{\partial X^{\mu}}{\partial \sigma}.
\eeq
Lars was fascinated by the enormous impact that Nambu had on the early foundations of string theory and was later instrumental in putting together 
  a memorial volume to honour Nambu\footnote{ {\it Memorial Volume for Y. Nambu}, Eds. L. Brink, L.N. Chang, M.-Y. Han and K.K. Phua, World Scientific (2016).}. 
  In 2008 Nambu was awarded the Nobel prize for something unrelated to string theory,
namely broken symmetries and the Nambu-Goldstone boson, although he practically single-handedly invented string theory as described above. The factorisation of the Virasoro amplitude
was within a year after Nambu also obtained  by L. Susskind and independently by H.B. Nielsen. The action principle for the string was soon after Nambu constructed also by Goto and is 
today known as the Nambu-Goto action. Neither of these  two key results by Nambu were  published and only appeared in talks and lecture notes (APS meeting in Chicago, January 1970, 
summer school in Copenhagen 1970) and an  unpublished manuscript
 (for the 1970 Rochester conference in Kiev)\footnote{See Nambu's own contribution to his collected works,  {\it Nambu, A Foreteller of Modern Physics}, Eds. T. Eguchi and M.Y. Han, World Scientific Series in 20th Century Physics, Vol 43 (2014).}.  More details about the life of Y. Nambu and his scientific achievements can be found in the book that Lars coauthored with 
 Pierre Ramond\footnote{Lars Brink and Pierre Ramond, {\it The Essence of a Genius;
A Tribute to Yoichiro Nambu}, World Scientific Series in 20th Century Physics: Volume 46, June 2023.}.

One of Lars' early papers that had a huge impact on the understanding of the string concerned  the physical origin of the critical dimension and the intercept, $D=26$ and $a=-1$. These facts, which 
actually contradict the hadron interpretation, 
were derived prior to Lars' work together with Holger 
Bech Nielsen\footnote{L. Brink and H.B. Nielsen, {\it A simple physical interpretation of the critical dimension of space-time in dual models}, Phys. Lett. {\bf 45B} (1973) 332.} 
but they showed that  $D=26$ and $a=-1$ could be related to the zero point energy of the bosonic string with similar results for other string models. In fact, by performing a physical
regularisation of the naively infinite zero point energy (involving the infinite sum over all positive integers) somewhat similar to the first calculations of the Casimir energy,
 Lars and Holger  arrived at the following result in $D$ spacetime dimensions:
\beq
a=-\frac{D-2}{24}.
\eeq

An interesting episode related to this is that at a summer meeting in Aspen in 1974 Lars was asked by Nambu to visit him in his office. Lars thought it was to discuss the latest developments in dual string 
theory but, in fact, it was to congratulate him on the beautiful paper he had written with Holger Bech Nielsen on getting the critical dimension from  summing up the zero point fluctuations of the string. 
The standard attitude today is to
attribute the fact that 
\beq
1+2+3+4+...=-\frac{1}{12},
\eeq
to a classic result in the theory of analytic functions: It follows directly  from  the Riemann zeta function $\zeta(s)=\Sigma_{n=1}^{\infty}\frac{1}{n^s},\,Re(s)>1,$ 
analytically continued and evaluated at $s=-1$. 
As Lars repeatedly pointed out
to his friends, the physics community was largely unaware of this mathematical result in 1973 which made his paper with Holger even more extraordinary.
Today this fact
is an absolutely fundamental insight for anyone with an interest in string theory and can be found in all textbooks on the subject.

Another kind of basic contributions by Lars in the year of 1973  appeared  in several papers together with David Olive, and in a couple of cases also with J. Scherk and C. Rebbi, on the construction
of unitary string loop diagrams. From previous works by Brower and by Thorn it was known that the Veneziano amplitude was free of negative norm states (ghosts) provided $D<26$ and the intercept $a=-1$.
The problem was to get only physical modes to propagate also in the loop diagrams which requires the insertion of projection operators\footnote{L. Brink and D. Olive, {\it The physical state projection operators
in dual resonance models for the critical dimension of space-time}, Nucl. Phys. {\bf B56} (1973), 253.}.
This step (for $D=10$) was particularly complicated in the case of the fermionic dual string models. Here one also  needed to know the proper set of gauge conditions which Lars investigated 
in a paper together with D. Olive, C. Rebbi and 
J. Scherk\footnote{L. Brink, D. Olive, C. Rebbi and J. Scherk, {\it The missing gauge conditions for the dual fermion emission vertex   and their consequences.}, Phys. Lett. {\bf 45B} (1973), 379.}.
In order to solve  these problems Lars and his collaborators used Feynman's tree theorem to define loop amplitudes  by sewing tree graphs without propagating ghosts. 
Lars often mentioned the {\it tree theorem}  as an example of the greatness of Feynman\footnote{Lars kept for his whole life  a note on his office wall where Feynman thanks him for using the tree theorem.}. 

Up to this point the papers by Lars mentioned above were all part of his doctoral thesis defended, as mentioned already,  in December of 1973.

The next set of key contributions by Lars and his collaborators took place in 1976 which turned out to be a very interesting year with several parallel developments involving local supersymmetry.
As already mentioned, in 1974 Yoneya, followed by  Scherk and Schwarz, developed the non-hadronic, that is the Yang-Mills and gravitational picture, of the open and closed string models.
In order to understand the physics of string theory (in particular spacetime fermions) it became crucial to develop a fermionic version of the bosonic dual string. 
The necessary ingredients had been discovered already in 1971, first by P. Ramond  and for a different sector of
the theory by  A. Neveu and J. Schwarz. Together these constitute what we today  know as the  {\it RNS} superstring. The  
fact that modern string theory could be turned into a success story relied 
heavily on this development and the fact that the tachyonic particles that appear can be seen to decouple and thus eliminated from the theory as first shown in 1977 by 
F. Gliozzi, J. Scherk and D.I. Olive\footnote{F. Gliozzi, J. Scherk and D.I. Olive, {\it Supergravity, Supergravity Theories and the Dual Spinor Model}, Nucl. Phys. {\bf B122} (1977) 253.}. 
The problematic nature of the tachyon is discussed at length at the end of  Lars' 
doctoral thesis. In fact, on page 45 in his thesis Lars is making the remarkable prediction that the elimination of the unphysical tachyon should happen through a {\it decoupling} instead
of the construction of a new tachyon-free dual string model.

The existence of the fermionic dual string models called for a supersymmetric action principle in $1+1$ dimensions, that is on the worldsheet of the string. 
To find such an action Lars  first developed (submitted to Physics Letters in August of 1976 and thus prior to the discovery of spacetime supergravity (see below)) the simpler case of a locally supersymmetric 
relativistic particle in  collaboration with also  Stanley Deser, Bruno Zumino, Paolo Di Vecchia and 
Paul Howe\footnote{L. Brink, S. Deser, B. Zumino, P. Di Vecchia and 
P. Howe, {\it The local symmetry of spinning particles}, Phys. Lett. {\bf B64} (1976) 435 - 438.}. The fermionic  string, or spinning string,   action was obtained the month after this
and demonstrated to be locally supersymmetric in $1+1$ dimensions, in a classic paper together with Paolo Di Vecchia and Paul Howe. The action they found
\footnote{L. Brink, P. Di Vecchia and P. Howe, {\it A locally Supersymmetric and Reparametrization Invariant Action for the Spinning String}, Phys. Lett. {\bf 65B} (1976) 471 - 474.}
 is written in terms of 
the string coordinates $X^{\mu}(\tau, \sigma)$ and their anticommuting (world-sheet) spinorial partners $\psi^{\mu}(\tau, \sigma)$. It reads
\beq
\label{spinningstringaction}
S[X^{\mu}, \psi^{\mu}; e_{a}{}^{\alpha}, \chi_a ]=\int d\tau d\sigma\,e(-h^{\alpha\beta}\half\partial_{\alpha} X^{\mu}\partial_{\beta} X^{\nu} \eta_{\mu\nu}-
\frac{i}{2}\bar\psi^{\mu}\gamma^{\alpha}\partial_{\alpha}\psi^{\nu}\eta_{\mu\nu}\nonumber
\eeq
\beq
\frac{i}{2}\bar\chi_{\alpha}\gamma^{\beta}\gamma^{\alpha}\psi^{\mu}\partial_{\beta}X^{\nu}\eta_{\mu\nu}+
\frac{1}{8}(\bar\chi_{\alpha}\gamma^{\beta}\gamma^{\alpha}\psi^{\mu})(\bar\chi_{\beta}\psi^{\nu}\eta_{\mu\nu})).
\eeq
Here $e_{a}{}^{\alpha}(\tau, \sigma)$, $h_{\alpha\beta}$ and $\chi_a(\tau, \sigma)$ are the 2-dimensional frame field (zwei-bein), metric   and the Rarita-Schwinger field, respectively, making this action a 2-dimensional 
supergravity theory\footnote{Local supersymmetry is absolutely necessary since without it one can not eliminate the negative normed states coming from the operators in the expansion of $\psi^0$,
 the time component of $\psi^{\mu}$.}. 
Note that the Nambu-Goto action is here replaced by an ordinary Klein-Gordon like kinetic term which seems to have been  first done in the earlier paper in 1976 mentioned above  involving also S. Deser and B. Zumino. 
This construction generalised
to the supersymmetric case the fact that  the Klein-Gordon type action given by the first term in the  above
$S[X^{\mu}, \psi^{\mu}; e_{a}{}^{\alpha}, \chi_a ]$  is classically equivalent on-shell to  the Nambu-Goto action. The paper by Brink, Di Vecchia and Howe was submitted to Physics Letters B 
in September 1976, just a few months after the breakthrough that had taken place concerning the construction of D=4 supergravity (with one supersymmetry) by 
Deser and Zumino,
and independently by  Ferrara, Freedman, and van Nieuwenhuizen\footnote{See the historical note by S. Deser on these developments arXiv: 1704.05886v3 [physics.hist-ph].}. 
That  string theory given by this  action  gives rise to a low energy effective field  theory in 10 spacetime dimensions that  is also supersymmetric (in spacetime) was
known from arguments presented  in September 1976 by Gliozzi, Scherk and Olive\footnote{F. Gliozzi, J. Scherk and D. Olive, {\it Supergravity and the spinor dual model}, Phys. Lett. {\bf 65B} (1976) 282.}.
The actions of the kind given in (\ref{spinningstringaction}) are today usually referred to as Polyakov actions as a result of his papers presenting the quantum version from a path integral 
perspective\footnote{A.M. Polyakov, {\it Quantum geometry of bosonic strings}, Phys. Lett. {\bf 103B} (1981), 207 - 210 and {\it Quantum geometry of fermionic strings}, Phys. Lett. {\bf 103B}(1981) , 211 - 213.}.

Also in 1976 Lars, together with what he referred to as an Italian national football team, namely 11 or 12 Italians, generalised in three papers the fermionic spinning string models of Ramond and Neveu-Schwarz
by adding new degrees of freedom in an attempt to reduce the critical dimension to $D=4$. This failed but the models constructed, having additional world-sheet local supersymmetries, increased the understanding of string theory
in general and are heavily 
cited\footnote{M. Ademollo et al, {\it Supersymmetric strings and colour confinement}, Phys. Lett. B{\bf 62}, 105 (1976), {\it Dual string with U(1) colour symmetry} Nucl. Phys. B{\bf 111}, 77 (1976), 
{\it Dual models with non-abelian colour and flavour symmetries} Nucl. Phys. B{\bf 114}, 297 (1976).}. This was probably one of the  last attempts to find a hadronic string theory.

Today string theory has created entirely new avenues towards this goal, now via holography and AdS/CFT related methods. Lars seemed
to the end of his life to stick to the old beliefs that the critical dimension could one day be reduced to four and thus that a directly physical dual string theory could be found.

Remarkably, Lars most cited paper is not one of the previously mentioned ones but instead  another paper  written in 1976 and submitted  to 
Nuclear Physics  in December\footnote{L. Brink, J.H. Schwarz and J. Scherk, {\it Supersymmetric Yang-Mills Theories}, Nucl. Phys. {\bf B121} (1977) 77.}. 
 It concerned the construction of maximally supersymmetric Yang-Mills theories in various dimensions, more specifically\footnote{They did not seem to be interested
in three dimensions and hence missed that interesting case.}  2,4,6 and 10. One important point made in the paper is that the $D=10$ theory is very simple. The action is
\beq
S_{D=10}=\int d^{10}x\left( -\frac{1}{4}F^a_{\mu\nu}F^{a\mu\nu}+i\bar\lambda^a\gamma^{\mu}D_{\mu}\lambda^a \right),
\eeq
where $D_{\mu}$ is the Yang-Mills covariant derivative in the adjoint irrep of the gauge group. This theory  is supersymmetric, but only on-shell,  in $D=10$ for Majorana-Weyl  spinors $\lambda$ due to the Fierz identity
\beq
f^{abc}(\bar\lambda^a\gamma^{\mu}\lambda^b)\bar\lambda^c\gamma_{\mu}=0.
\eeq
The second key result is that by dimensional reductions from $D=10$ to the lower dimensions listed above one straightforwardly obtains the maximally supersymmetric Yang-Mills theories 
in these dimensions which are a bit more complicated. This paper seems to have been the start for Lars'  life-long interest in these theories. 
The theory in $D=4$ space-time dimensions has four supersymmetries and its action reads (with $i=1,2,3,4$)
\beqa
S_{D=4}&=&\int d^4x( -\frac{1}{4}F^a_{\mu\nu}F^{a\mu\nu}+i\bar\chi^{ai}\gamma^{\mu}D_{\mu}\chi^{ai}+\half D^{\mu}\Phi^a_{ij}D_{\mu}\Phi^a_{ij}\cr
&-&\frac{i}{2}g\,f^{abc}(\bar\chi^{ai}\chi^{jb}\Phi^c_{ij}) -\frac{1}{4}g^2f^{abc}f^{ade}\Phi^b_{bij}\Phi^c_{kl}\Phi^{dij}\Phi^{ekl}).
\eeqa
We will have reason to return to this theory shortly due to a speculation in a paper in 1982 where Lars, together with M. Green and J.H. Schwarz, derived  the maximally 
supersymmetric Yang-Mills and gravity theories in $D=4$ as limits of the superstring. The speculation concerned the fact that since the above super-Yang-Mills theory is obtained this way  it might 
actually be finite  to all orders in perturbation theory\footnote{L. Brink, M.B. Green and J.H. Schwarz, {\it $\mathcal N=4$ Yang-Mills and $\mathcal N=8$ supergravity as limits of string theories.}, 
Nucl. Phys. {\bf B198} (1982) 474. It seems, according to Lars, that this speculation was first articulated by M. Gell-Mann some
time before  this paper appeared.}. 

While local supersymmetry was necessary and directly used as a tool in the 1+1 dimensional theory it took some time before 
F. Gliozzi, D. Olive and J. Scherk
realised that  the theory in spacetime, or rather its low energy effective field theory, must also exhibit supersymmetry as already mentioned above. This led to a race to find the
Lagrangian which describes the interacting theory of the graviton and the spin 3/2 field, the so called Rarita-Schwinger field. These rather complicated results were obtained 
about the same time by two different groups, Deser and Zumino and independently by Ferrara, Freedman and van Nieuwenhuizen (see above). 
The rather complicated approaches used by these groups called for more powerful techniques.    
One such was superspace which was developed for supergravity by Wess and Zumino in 1976. The paper was submitted to Physics Letters the day after Brink, Schwarz and Scherk submitted their paper on supersymmetric Yang-Mills theories to Nuclear Physics at the very end of the year.  This marked the end of a truly remarkable year  for dual string models and supersymmetry, and for Lars with his many
fundamental contributions to these developments. 

We may remark here that Lars, together with J.O. Winnberg, developed a worldsheet superspace formalism in a paper submitted already in 
November 1975\footnote{L. Brink and J.O. Winnberg, {\it The superoperator formalism of the Neveu-Schwarz-Ramond model}, Nucl. Phys. {\bf B103} (1976)445.}, while Lars, now together with John Schwarz,
a few years later studied other features of superspace and theta variables in their rather  well-cited paper 
on {\it Quantum superspace}\footnote{L. Brink and J.H.  Schwarz, {\it Quantum superspace}, Phys. Lett. {\bf 100B} (1981) 310.}.

An alternative   description of the superstring, but fully equivalent to the $NSR$ model, was later developed by Green and Schwarz, and here Lars also made important contributions during his 
frequent visits to Caltech. This new formulation used  to study the superstring is very close to what is known as superspace in field theory, namely the superspace coordinates
$x^{\mu}, \theta^m$ in supergravity field theory are turned into world-sheet fields $(X^{\mu}(\tau,\sigma), \Theta^m(\tau,\sigma))$ providing an alternative to the RNS formulation which
contains a fermionic field which is a space-time vector ($\psi^{\mu}$) instead of the fermionic world-sheet scalar which is  a space-time spinor ($\Theta^m$).
Lars had a keen interest also in the superspace description of both Yang-Mills and supergravity. In three papers with M. Gell-Mann, P. Ramond and J.H. Schwarz Lars studied the geometric
approach to supergravity\footnote{See, e.g.,  M. Gell-Mann, P. Ramond and J. Schwarz, {\it Supergravity as geometry of superspace},  Phys. Lett. {\bf B74} (1978) 336 - 340.}. This paved the way for two nice papers together with P. Howe, then a postdoc in Lars' group at Chalmers, in 1979 and 
1980\footnote{L. Brink and P. Howe, {\it $\mathcal N=8$ supergravity in superspace}, Phys. Lett. {\bf B74} (1979) 268, and {\it Eleven-dimensional supergravity on the mass-shell in superspace}, 
Phys. Lett. {\bf B91} (1980) 384.}. In these two papers Lars and Paul rederived the maximally supersymmetric gravity theories in $D=4$ and $D=11$ dimensions from the superspace 
Bianchi identities by imposing a suitable set of constraints on the super-torsion tensor.

In the 1980s Lars continued, together with his students and collaborators at Chalmers,  to work on issues which grew out of questions and techniques he
had studied earlier in the string theory context. One such technique was the light-cone approach to strings and field theories. A theory that he contributed to in several papers was
maximally supersymmetric Yang-Mills theory which he regarded as his "baby", to quote his own words from an interview with him for American Institute of Physics in
 May 2021\footnote{ This interview can be found on the English version of the Wikipedia site for Lars. }.
The study of this theory started during a visit to Caltech in 1976 with a paper with John Schwarz and 
Joel Scherk  where the theory was constructed in dimensions $D=10$ as already mentioned above.
They also demonstrated how one can relate the theories in different dimensions to each other by dimensional reduction. Another  important result in this context was obtained in  1982 when Lars
visited Caltech and showed in a paper with J.H. Schwarz and M. Green how to, in some limits of string theory, obtain $\mathcal N=8$ supergravity as well as $\mathcal N=4$ super-Yang-Mills (see above).

Shortly after this Lars, then  back at Chalmers, together with
Olof Lindgren and Bengt E.W. Nilsson, applied supergraph techniques in the light-cone gauge to this $\mathcal N=4$ super-Yang-Mills theory and proved it to be 
 a quantum mechanically  finite  theory, the  first one of this kind and hence a result of fundamental importance for the much of the formal high energy developments in the following years. 
The proof was crucially based on the light-cone 
superfield\footnote{See L. Brink, O. Lindgren and B.E.W. Nilsson, {\it N=4 Yang-Mills Theory on the Light-Cone}, Nucl. Phys. B{\bf 212}, 401 (1983).} which collects all the independent physical degrees 
of freedom\footnote{Two photon states, six scalars and four complex spin 1/2 states. Note that $i=1,2,3,4$ and that $\Phi_{ij}(x)$ is complex self-dual representing six real scalar fields.}  
of the theory in one object
\beq
\Phi^a(x, \theta)=\frac{1}{\partial^+}A^a(x)+\frac{i}{\partial^+}\theta^i\lambda^a_i(x)+\frac{i}{\sqrt 2} \theta^i\theta^j \Phi^a_{ij}(x)+
\frac{\sqrt 2}{6} \theta^i\theta^j\theta^k \epsilon _{ijkl}\bar\lambda^{ak}(x)+\frac{1}{12}\partial^+(\theta)^4\bar A^a(x),
\eeq
where all fields are in the adjoint irrep of the gauge group as indicated by $a$. That such a (complex) superfield exists relies on two facts: 1) 
 The momentum light-cone component $p^+\equiv \half(p^0+p^3)$, above written as $\partial^+$,  is declared positive definite, and 2) 
 the light-cone gauge, $A^+\equiv \half(A^0+A^3)=0$,  that makes it possible to explicitly solve for $A^-$ in terms of the two photon polarisations (here in $A(x)\equiv A^1+iA^2$). 
In Yang-Mills loop-calculations, using so called supergraphs,  this superfield gives rise to  combinations of a huge numbers of ordinary Feynman diagrams resulting in loop-expressions
which can then be shown to be finite at any 
loop-order\footnote{See L. Brink, O. Lindgren and B.E.W. Nilsson, {\it The Ultraviolet Finiteness of the N=4 Yang-Mills Theory}, Phys. Lett. B{\bf 123}, 323 (1983). 
A similar proof of finiteness was presented by S. Mandelstam about the same time.}. More historical details on these developments can be found in Lars' own 2015 account of the 
events\footnote{L. Brink, {\it Maximally Supersymmetric Yang-Mills Theory, The Story of ${\mathcal N}=4$ Yang-Mills Theory}, arXiv 1511.0297 [hep-th].}.
 Lars often emphasised the enormous importance of this
result for later developments in string theory, not least for the emergence of the AdS/CFT correspondence at the end of the 1990s.

Soon after this important development Lars applied these light-cone methods to higher spin theories which turned out to be a tremendous success.
The construction of higher spin three-point vertices in a flat space-time together with two of his 
PhD students\footnote{A.K.H. Bengtsson, I. Bengtsson and L. Brink, {\it Cubic interaction terms for arbitrary spin}, Nucl. Phys. {\bf B227} (1983) 31 - 40 and 
 {\it Cubic interaction terms for arbitrarily extended supermultiplets}, Nucl. Phys. {\bf B227} (1983) 41 - 49.}
 started an active subarea of field theory that has been thriving ever since. More recently Lars tried to use these light-cone methods
to argue that  supergravity with the maximal number of  supersymmetries, that is eight,  for which a superfield  similar to the one used in the super-Yang-Mills theory can be constructed,  
is also a finite quantum field theory.  Despite the collective effort of many people in the field 
to show this it is still far from being proven. Lars probably nourished the idea that maximally supergravity in D=4 is finite for the rest of his life. In fact, he later introduced
some new techniques based on the scalar field coset structure in these theories to study  possible restrictions on  counter terms and possible implications on  the issue  of
 finiteness\footnote{See, e.g., L. Brink, S.S. Kim and P. Ramond, {\it E7(7) on the light cone}, JHEP 2008 (06), 034.}.

When Yoneya, and independently  Scherk and Schwarz,  in 1974 studied the consequences of  the massless spin-2 particle in the spectrum  interpreting it as a graviton  it was clear that the Regge slope parameter
should be connected to the Planck scale instead of being 1 $(GeV)^{-2}$ as dictated by hadronic physics.  This meant that modern string theory was born. String theory not only generated
from basic principles (that is quantum mechanics in 1+1 dimensions) a theory of spacetime gravity but it also solved one of the most profound problems in our description of the Universe,
the non-renormalizability issue of Einstein's theory of gravity based on the Einstein-Hilbert action principle. Lars, together with most researchers in his generation, saw this
as a hint of a TOE, a theory of everything.  In particular this implied that string theory should be capable of predicting a unique vacuum and hence provide a full fundamental explanation
of the standard model including  particle masses etc.
This attitude towards the role of string theory has largely been abandoned in recent years in favour of the more formalistic point of view similar to the one we have towards ordinary quantum mechanics.
When string theory, and in particular the spin-off  AdS/CFT, eventually got applied to everything from nuclear physics to  superconductivity most people in the field shifted attitude towards the new more formal one. Lars
eventually came to the conclusion that this shift in attitude was  necessary but he probably 
never really came to fully embrace it\footnote{A most personal reflection by one of the authors (BEWN).}.

To make clear that Lars had a broader interest in physics than just the field of high energy physics and string theory, we  mention here the   important paper by Lars
on the Calogero model where he and his collaborators managed to find an explicit solution to the old N-body Calogero 
problem\footnote{L. Brink, T.H. Hansson and M.A. Vasiliev, {\it Explicit solution to the  N-body Calogero problem}, Phys. Lett. {\bf B286}, 109.}. 

 \section{Lars Brink, some final reflections}

 People close to Lars were often struck by his fantastic memory. He sometimes took advantage of this for instance when he recited long poems by
 authors he had a particular relation to, like Nils Ferlin. Besides the keen interest in humanities that he demonstrated in this and other ways he was also 
 a huge fan of athletics, soccer and, in fact, sports in general. It could  be quite intimidating to realise that he remembered an enormous number of results by athletes active 
 decades ago and whose names were long forgotten. A memory related to himself that he was particularly fond of was  his visit, in 1958, to the newly built
 soccer stadium in G\"oteborg to watch a soccer game in the then on-going world 
 championship in Sweden. It was probably one of the semi-finals. 
 From other sources (his family) we later learned that he was actually selling candy at the stadium, a fact that he apparently ``forgot''  to inform 
 his friends about. 
 His memory bank also contained an abundance of anecdotes,
 often of a humorous kind, and frequently  involving 
 various well-known people in physics like Murray Gell-Mann and many many others. Although recalled in a humorous tone these anecdotes
 sometimes expressed in addition the enormous admiration he felt for many of the giants in high energy physics.

With a slight smile on his face Lars could occasionally  comment on the fact that he did not get very far in life. What he had in mind was the  fact that he grew up in Johanneberg
in central G\"oteborg and went to high school down the street and to university across the same street where he also spent his whole academic career, 
neither of these schools located more than a kilometer away from his home.

After having been ill for a long time  Lars passed away peacefully  in October 2022. Despite being confined to his bed at  home he was active giving talks via  
internet\footnote{ Via internet Lars honoured the achievements of Madame Wu in a talk at the C.S. Wu Global Online Symposium, September 24th, 2022.}and writing articles 
  until the very last weeks of his life\footnote{Lars'  last paper appeared on September 2, 2022, and was a final tribute to Y. Nambu: L. Brink and 
  P. Ramond, {\it Nambu and the Ising Model}, arXiv 2209.01122[hep-th].}. We have lost a very good friend, a most stimulating 
  colleague and someone who fought fiercely for  curiosity-driven fundamental research of all kinds.

\section*{Acknowledgement}
We are very grateful to Paolo Di Vecchia, Marc Henneaux and Pierre Ramond for a thorough reading of the manuscript and for a number of 
suggestions which have   improved the presentation.


\end{document}